\documentclass[showpacs,showkeys,preprintnumbers,amsmath,amssymb]{revtex4}

\usepackage{amsmath}
\usepackage{amsfonts}
\usepackage{amssymb}
\usepackage[colorlinks,linkcolor=blue,citecolor=red]{hyperref}
\usepackage{graphicx}
\usepackage{subfigure}
\usepackage{graphicx}
\usepackage{dcolumn}
\usepackage{bm}

\def\be{\begin{equation}}
 \def\ee{\end{equation}}
\def\bea{\begin{eqnarray}}
\def\eea{\end{eqnarray}}

\begin{document}


\title{Comments on $^"$Cosmic evolution in Brans-Dicke chameleon cosmology$^"$}

\author {Haidar Sheikhahmadi}
\email{h.sh.ahmadi@gmail.com; h.sheikhahmadi@ipm.ir }
\affiliation{School of Astronomy, Institute for Research in Fundamental Sciences (IPM),
 P. O. Box 19395-5531, Tehran, Iran,}
\affiliation{Institute for Advanced Studies in Basic Sciences (IASBS) Gava Zang, Zanjan
45137-66731,
Iran}



\date{\today}

\begin{abstract}
The authors of Ref. \cite{1-2}, investigated cosmic evolution in an external interacting model of scalar tensor gravity namely Brans Dicke chameleon scenario. The procedure of this work  contains novelties  but, it shall be observed from this comment, their report faces three fundamental drawbacks. One of them concerns the energy conservation equation, and the other two flaws are about mathematical mistakes. In the scalar tensor gravity models, a well-known method, in order to obtain conservation equation one must combine  Friedmann equations with modified Klein-Gordon equation. But in  Ref. \cite{1-2}, by virtue of the mentioned approach one would not be able to achieve a correct result for conservation equation. In addition, one can readily realize that their mathematical mistakes lead to different results compared to the present report.
\end{abstract}

\keywords{\textcolor[rgb]{0.00,0.50,0.00}{Chameleon Brans Dicke gravity, Modified conservation equation}}
\maketitle

\newpage

\textcolor[rgb]{0.00,0.00,1.00}{\section{Introduction}}

As an interesting scalar tensor model, one can refer to the well known Brans Dicke (BD) model of gravity {\cite{2}}. To avoid any repetitive discussions  about BD model and its applications, which could be found in  a wide range of literature, we briefly mention only some drawbacks of  BD scenarios in the following. In fact, more studies have shown that, BD theory faces some problems in comparison to observation. For instance although BD
theory proved to be useful for the solution to many cosmological
problems, it has a serious drawback through which the BD parameter $\omega$, has to have a small value of order unity. Also, it is realized from recent literature within past two decades, a suitable framework to overcome mentioned problems and investigate the evolution of the Universe, from the primordial era to late time is chameleon mechanism {\cite{1}. In this approach, Newtonian constant of gravitation $G$ is constant but scalar field has non-minimal coupling with Lagrangian of matter which leads to describing the Universe evolution in a better way rather than BD model.
In fact, one of the main ideas that caused to introduce the chameleonic behaviour goes back to providing an alternative mechanism for circumventing the constraints  on local test of gravity, especially the fifth force concept {\cite{1a}}. In the chameleon model of gravity the scalar field
acquires a mass whose magnitude depends on the local matter density, and so gives an effective mass to
a light scalar field via field self-interaction and interaction between the scalar field and matter{\cite{1}.\\
To solve some shortages that chameleon model has faced to them, some researchers, for example Clifton \textit{et al}.,
{\cite{3}} and Das \textit{et al}., {\cite{4}} have proposed a
framework in which scalar field has nonminimal coupling with both the
geometry and matter,  i.e., BD chameleon (BDC) mechanism. The BDC model is useful  even for high values of $\omega$ and thus it
 is in good agreement with the observational data {\cite{5}}. Also we can emphasize that, by virtue of these scalar tensor models of gravity, different  scenarios, for instance holographic {\cite{6}} and new agegraphic {\cite{7}} models of dark energy (DE), have been investigated. For more studies in each areas we can refer the reader to {\cite{7a, 8}}.\\
Based on discussions above, it will be shown the  Eqs. (2), (7), (9), and therefore related results of Ref. \cite{1-2} are wrong.


\textcolor[rgb]{0.00,0.00,1.00}{\section{Mathematical toolkits for BDC scenario}}
Following Ref. \cite{1-2}, we want to consider the action
\begin{equation}\label{4}
A=\int d^4x\sqrt{-g}\left(\phi R-
\frac{\omega}{\phi}\partial_{\mu}\phi\partial^{\mu}\phi-V(\phi)+2f(\phi)\mathcal{L}_{m}\right),
\end{equation}
where in the last term we notice the factor $2$  which is absent in Ref. \cite{1-2}, and in the following we shall see this factor $2$ will lead to quite different results. In above equation, $R$ is the Ricci scalar,  $\omega$ is the dimensionless BDC parameter,
$\mathcal{L}_{m}$ is the Lagrangian of matter, and $\phi$ is the BDC
scalar field with a potential $V(\phi)$. The last term in the action indicates the interaction
between  matter lagrangian and some arbitrary function $f(\phi)$
of the BDC scalar field, and we shall assume $f(\phi)\neq1$, throughout this work. It should be noticed that we  shall consider homogeneous and isotropic Fridmann-Limature-Robertson-Walker (FLRW) metric  as
\begin{equation}\label{5}
ds^{2}=-dt^{2}+a^{2}(t)\left[\frac{dr^{2}}{1-kr^{2}}+r^{2}(d\theta^{2}+\sin^{2}\theta
d\phi^{2})\right],
\end{equation}
where $a(t)$ refers to the scale factor and $k=-1, 0, +1$ indicates open,
flat and close Universe, respectively. Following the approach in Ref. \cite{1-2}, but using usual definition of energy momentum tensor $(T_{\mu\nu} = \frac{-2}{\sqrt{-g}} \frac{\delta (\sqrt{-g}\mathcal{L}_m)}{\delta g^{\mu\nu}})$, by  varying the action with respect to
$g_{\mu\nu}$ the field equations could be obtained as follows
\begin{equation}\label{5a}
\phi(R_{\mu \nu}-\frac{1}{2}g_{\mu \nu}R)=  f(\phi)T_{\mu
\nu}+\frac{\omega}{\phi}(\partial_{\mu} \phi \partial_{\nu} \phi
- \frac{1}{2}g_{\mu \nu}(\partial_{\alpha}\phi)^2)
+ [\nabla_{\mu} \nabla_{\nu} -g_{\mu \nu}\Box]\phi -
g_{\mu\nu}\frac{V(\phi)}{2}.
\end{equation}
Considering $0-0$ and $i-i$ the components of Eq.(\ref{5a}), one gets
\begin{equation}\label{6}
3\left(\frac{\dot{a}^{2}}{a^{2}}+\frac{k}{a^{2}}\right)=\frac{f(\phi)}{\phi}\rho-
3\frac{\dot{a}}{a}(\frac{\dot{\phi}}{\phi})
+\frac{\omega}{2}\frac{\dot{\phi}^{2}}{\phi^{2}}+\frac{V(\phi)}{2\phi},
\end{equation}
\begin{equation}\label{7}
2\frac{\ddot{a}}{a}+\frac{\dot{a}^{2}}{a^{2}}+\frac{k}{a^{2}}=-\frac{f(\phi)}{\phi}p
-\frac{\omega}{2}\frac{\dot{\phi}^{2}}{\phi^{2}}-2\frac{\dot{a}}{a}(\frac{\dot{\phi}}{\phi})
-\frac{\ddot{\phi}}{\phi}+\frac{V(\phi)}{2\phi}.
\end{equation}
Here ${\dot{a}}/{a}$ is the Hubble parameter and overdot indicates
differentiation with respect to the cosmic time $t$. Also $\rho$ and $p$
refer to the  energy density and  pressure, respectively.\\
In addition, the Klein-Gordon equation could be readily achieved as
\begin{equation}\label{8}
\Box\phi=\frac{1}{2\omega+3}\left(f T-{2}{\cal L}_m  \phi f_{,\phi}\right)
+\frac{1}{2\omega+3}\left(\phi V_{,\phi}-2V\right).
\end{equation}
As indicated in Refs. {\cite{8b, 8a}}, when one has the interaction between Lagrangian of matter and scalar field the best choice for ${\cal L}_m $ is pressure $ p$. But whereas our main aims, in writing this report, are about  the mathematical and obvious problems so following Ref. \cite{1-2} we consider ${\cal L}_m = T/4$ in which $T$ is the  trace of energy-momentum tensor.
Here, we suppose that all components of matter are perfect fluid  and therefore, the  stress-energy tensor can be expressed as
\begin{equation}\label{9}
T_{\mu\nu}=(\rho+p)u_{\mu}u_{\nu}+pg_{\mu\nu},
\end{equation}
where $u^{\mu}$ is the four-vector velocity of the fluid. By
varying the action with respect to the BDC scalar-field, and by virtue of Eq. (\ref{5}) the dynamical
equation of $\phi$ is obtained as follows
\begin{equation}\label{10a}
\ddot{\phi}+3H\dot{\phi}=\frac{\rho-3p}{2\omega+3}\left(f-\frac{1}{2}\phi
f_{,\phi}\right)-\frac{2}{2\omega+3}\left(V-\frac{1}{2}\phi
V_{,\phi}\right).
\end{equation}
Combining Eqs. (\ref{6}, \ref{7}) and (\ref{10a}) we get the modified conservation equation as
\begin{equation}\label{17}
\dot{\rho}+3H\rho(1+\omega_{t})=
\frac{-3\dot{f}}{4f}(1+\omega_{t})\rho,
\end{equation}
where $\omega_{t}=p/\rho$. Based on discussions in Ref. \cite{8a}, $p$ and $\rho$ indicate pressure and energy density for all components of the Universe except for scalar field.
According to Refs. \cite{8b, 8a}, if we consider  ${\cal L}_m = p$, the conservation equation of energy becomes
\begin{equation}\label{15}
\dot{\rho}+3H\rho(1+\omega_{t})=
-\frac{\dot{f}}{f}(1+\omega_{t})\rho.
\end{equation}
And again the right hand side of conservation equation is not equal to zero and therefore leads to different results compared to Ref. \cite{1-2}.\\
In next section it will be shown, in more details, in Ref. \cite{1-2} the right hand side of the above equation was neglected completely and it causes omitting the chameleonic effects in mentioned paper.

\textcolor[rgb]{0.00,0.00,1.00}{\section{Comments}}

\begin{itemize}
  \item[I,]\textbf{\textcolor[rgb]{0.00,0.50,1.00}{About the action}}\\
 The first step of every work in scalar tensor gravity is to determine the action. Then by taking the variation with respect to the independent parameters, i.e. the metric and scalar field, the main equations  can be obtained. In Ref. \cite{1-2}, the action considered is
\begin{equation}\label{1}
A=\int d^4x\sqrt{-g}\left(\phi R-\frac{\omega}{\phi}\partial_{\mu}\phi\partial^{\mu}\phi-V(\phi)+f(\phi)\mathcal{L}_{m}\right),
\end{equation}
hence taking the variation of the action with respect to the metric gives us the field equation as
\begin{equation}\label{2aa}
\phi(R_{\mu \nu}-\frac{1}{2}g_{\mu \nu}R)= \frac{1}{2} f(\phi)T_{\mu
\nu}+\frac{\omega}{\phi}(\partial_{\mu} \phi \partial_{\nu} \phi
- \frac{1}{2}g_{\mu \nu}(\partial_{\alpha}\phi)^2)
+ [\nabla_{\mu} \nabla_{\nu} -g_{\mu \nu}\Box]\phi -
g_{\mu\nu}\frac{V(\phi)}{2}.
\end{equation}
It should be noted that to get this result, we considered
$T_{\mu\nu} = \frac{-2}{\sqrt{-g}} \frac{\delta (\sqrt{-g}\mathcal{L}_m)}{\delta g^{\mu\nu}}$.
As can be seen  there is a  coefficient $1/2$ in  Eq.(\ref{2aa}) for $T_{\mu\nu}$ tensor, which was neglected in Ref. \cite{1-2}. It can be checked that the result of  \cite{1-2} for Eq. (2) is obtained if and only if they consider $T_{\mu\nu} = \frac{-1}{\sqrt{-g}} \frac{\delta (\sqrt{-g}\mathcal{L}_m)}{\delta g^{\mu\nu}}$, otherwise they should correct the form of the action. Now let us answer a big simple question: where does the definition of $T_{\mu\nu}$ come from?. To answer this basic question, we know that the form of the energy-momentum tensor directly comes from the action by taking variation with respect to the metric. So if the author of Ref. \cite{1-2} have used a different form of $T_{\mu\nu}$, they should have mentioned it in their work.\\
Now let us show in a brief calculation, how the coefficient $1/2$ appears in Eq.(\ref{2aa}) above? To do so, by virtue of Eq.(\ref{1}) one has
\[\delta \int {{d^4}} x\sqrt { - g} \,f(\phi ){{\cal L}_m} = \int {{d^4}} x\Big[\frac{{\partial (\sqrt { - g} \,f(\phi ){{\cal L}_m})}}{{\partial {g^{\mu \nu }}}}\delta {g^{\mu \nu }} + \frac{{\partial (\sqrt { - g} \,f(\phi ){{\cal L}_m})}}{{\partial {g_{,\alpha }}^{\mu \nu }}}\delta {g_{,\alpha }}^{\mu \nu }\Big]\]
\[=\int {{d^4}} x \big(\frac{{\partial (\sqrt { - g} \,f(\phi ){{\cal L}_m})}}{{\partial {g^{\mu \nu }}}} - \frac{\partial }{{\partial {x^\alpha }}}[\frac{{\partial (\sqrt { - g} \,f(\phi ){{\cal L}_m})}}{{\partial {g_{,\alpha }}^{\mu \nu }}}]\big) \delta {g^{\mu \nu }}\]
where a comma followed by an index indicates partial differentiation. By defining energy-momentum tensor as
\[{T_{\mu \nu }} = \frac{{ - 2}}{{\sqrt { - g} }}\{ \frac{{\partial (\sqrt { - g} \,{{\cal L}_m})}}{{\partial {g^{\mu \nu }}}} - \frac{\partial }{{\partial {x^\alpha }}}[\frac{{\partial (\sqrt { - g} \,{{\cal L}_m})}}{{\partial {g_{,\alpha }}^{\mu \nu }}}]\}, \]
we find
\[\delta \int {{d^4}} x\sqrt { - g} \,f(\phi ){{\cal L}_m} = \frac{{ - 1}}{2}\int {{d^4}} x\sqrt { - g} \,f(\phi )\,{T_{\mu \nu }}\delta {g^{\mu \nu }}.\]
Now by repeating such a procedure for other components of the action the results of Eq.(\ref{2aa}) will be obtained, which are different from Eq.(2) appeared in Ref. \cite{1-2}. Anyway, when we consider the usual form of energy-momentum tensor for their work, some additional terms on the right hand side of Eq.(\ref{17}) appear beside $\frac{-3\dot{f}}{4f}(1+\omega_{t})\rho$ . At last when we compare the results of  \cite{1-2} with the works in the literature it is obvious that by virtue of $T_{\mu\nu} = \frac{-2}{\sqrt{-g}} \frac{\delta (\sqrt{-g}\mathcal{L}_m)}{\delta g^{\mu\nu}}$, the action (\ref{4}) must be corrected as
\begin{equation}\label{4aa}
A=\int d^4x\sqrt{-g}\left(\phi R-
\frac{\omega}{\phi}\partial_{\mu}\phi\partial^{\mu}\phi-V(\phi)+2f(\phi)\mathcal{L}_{m}\right).
\end{equation}
For more details the reader is referred to \cite{7a, 8, 8b, 8a}.

  \item[II,]\textbf{\textcolor[rgb]{0.00,0.50,1.00}{About the second Friedmann equation}}\\

    Now, by comparing Eq.(\ref{7}) in this report with equation (7) in Ref. \cite{1-2}, it is realized that the coefficient $f(\phi)$ is completely neglected. It is obviously clear that in equation (2) of Ref. \cite{1-2}, $f(\phi)$ is the coefficient of energy-momentum tensor, so dimensionally $f(\phi)$ has to be appear as the coefficient of pressure $p$ in equation (7). In other words, as can be seen in Eqs. (6) and (7), the origin of energy density and pressure is  Lagrangian of matter. So from Eq.(2) of  \cite{1-2} it is obvious that, $f(\phi)$ has interaction with both energy density and  pressure. Therefore, if scalar field has interaction with energy density in Eq.(6), how the interaction was omitted in Eq.(7)?\\
   In fact, the importance of this question goes back to combination of Eqs.(6)-(8) in Ref. \cite{1-2} to get conservation equation, i.e. equation (9). Anyway with (or without) this coefficient, as discussed above, the right hand side of Eq.(9) in \cite{1-2} is not equal to zero. And in the following sub-section, we shall show that despite the title chosen for their report they have not taken into account the effects of chameleonic interactions.

\item[III,]\textbf{\textcolor[rgb]{0.00,0.50,1.00}{About the conservation equation}}\\

If one compares Eq.(\ref{17}) of this work and equations (9) in Ref. \cite{1-2} it is  obviously observed that the right hand side, which contains the effects of coefficient $f(\phi)$, is completely ignored. Hence the results of this comment and Ref. \cite{1-2} are completely different. In fact, based on their results, it is assumed that in BDC model each part of matter is able to justify the energy conservation separately. But as we have shown in this paper this is not true. In fact we should emphasize that although total energy is conserved but for each separate part  the conservation of energy  could not be justified individually, and it means that we have the energy flux between components of the Universe. For more details we refer the reader to Refs. \cite{8, 8a, 9}.
\end{itemize}

\textcolor[rgb]{0.00,0.00,1.00}{\section*{Acknowledgment}}
HS would like to thank the anynomous referee for his/her constructive and enlightening comments and for
his crucial clarifying points.
He also thanks B. Ahmadi for some useful suggestions.




\end{document}